# Evolution of structural ($\alpha$) relaxation-time anomalies in $Ge_xSe_{1-x}$ Chalcogenide glasses

Deepak Sharma*, Sujatha Sampath[#], and A.M. Awasthi[†]

*SCRIET, CCS University, Meerut- 250 004, India

[#]Department of Physics, Arizona State University, Tempe, AZ 85287-1504, USA

[†]UGC-DAE Consortium for Scientific Research, Indore- 452 001, India

*e-mail: deepak22phys@gmail.com

## ABSTRACT

We examine enthalpy relaxation across the chalcogenide glass series $Ge_xSe_{1-x}$, prepared over close-by compositions using conventional melt-quenching technique. We estimate the timescale $\tau(T_g)$ characterizing enthalpic relaxation near the kinetic glass transition temperature, using the non-reversing heat-flow data obtained from MDSC (modulated differential scanning calorimetry) measurements over a wide range of compositions ($2.1 \leq r \leq 2.8$, $r = 2x+2$). Anomalies in the enthalpy-relaxation characteristic-times $\tau_g(r)$ are identified as marking rigidity-transitions encountered in successive Ge-doping of polymeric selenium chains.

## 1. Introduction

J.C. Phillips first introduced the notion of highest glass forming propensity in covalent random networks (CRN) [1-2] near a mean atomic coordination number of $r = 2x+2 = 2.4$. Concurrently, Thorpe conceptualized the floppy and rigid vibrational modes [3] and the occurrence of 'rigidity threshold' in CRN's. Angell's subsequent classification of kinetically fragile/strong glass-formers [4] is in conformity with these ideas. In the glass transition studies using DSC [5-7], though elaborate thermal procedures are employed and their effects on the transition characteristics are reported, relatively scarce [8] relate them to appropriate physical parameters. The non-equilibrium nature [9] of temperature-scanning experiments has to be taken into account while interpreting their thermograms. Examining the non-reversing heat flow of $Ge_xSe_{1-x}$ covering a breadth of relaxation times is one way to explore these issues. The present work illustrates caloric and relaxation manifestations of the rigidity-transitions in a prototype glass-family Ge$_x$Se$_{1-x}$, established in several network-glasses via enthalpic and vibrational signatures, to facilitate rigorous theoretical modeling of the phenomenon. For investigations of the glassy nature, while the reversing heat flow (proportional to specific heat) clearly shows the glass transition range, the non-reversing heat flow is a measure of enthalpic relaxation, separately-obtained in the modulated differential scanning calorimetry (MDSC) experiments. The non-reversing heat flow has been analyzed to furnish the effective timescales governing the enthalpy relaxation rate.

## 2 Experimental

Bulk glasses of Ge$_x$Se$_{1-x}$ composition were synthesized from elemental Ge and Se (5N purity) using the standard melt-quench technique. Their XRD patterns showed no characteristic Bragg peaks [10] and the glassy nature of the samples has been reported [11-12]. To our knowledge, for the first time we reported [13] the confined acoustic vibrational modes in low-frequency Raman spectra of glassy Ge$_x$Se$_{1-x}$ samples. For the measurements, samples were encased in aluminium pans, with an empty pan as the reference. The MDSC measurements performed across $T_g$ at uniform heating rates of 5°C/min, combined with ±1.5°C per 80sec modulation, provided direct data for the reversing heat flow (proportional to heat capacity $C_p$) and for the non-reversing heat flow (proportional to the enthalpy relaxation rate $\dot{H}$) [14-15]. Study of the enthalpy relaxation data is presented here.

## 3 Results and Discussion

The non-reversing heat flow (NRHF) in an MDSC run is obtained as an endothermic peak vs. temperature (time). In order to get the sample characteristic an additive built-in instrumental offset needs to be removed from the raw data. For this purpose, time derivative of the raw data indicates the onset/offset points within which the intrinsic effect takes place. A smooth spline-fit to the background is then made, which also interpolates over the $T$-range of interest. This background fit is subtracted away from the raw data, to get the intrinsic kinetic signal, and one such processed sample-data (the intrinsic enthalpy-relaxation rate $\dot{H}$) is shown in fig.1. We define the peak temperature of $\dot{H}(T)$ as the *kinetic* glass transition temperature $T_k$. The glass transition temperature $T_g$ is generically taken where the magnitude of structural relaxation time is ~100 sec [16]. Presently, as per ref. [17], inflexion-point in the $C_p(T)$-step (fig.2a) is defined as the *caloric* glass transition temperature.

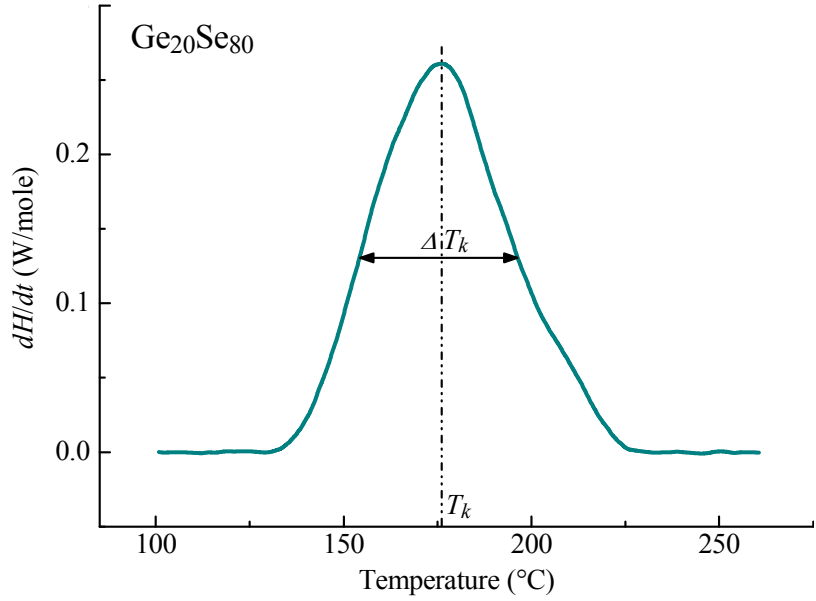

**Fig.1** Background-subtracted non-reversing heat flow (enthalpy-decay rate $\dot{H}$) across the *kinetic* glass transition temperature in $Ge_{20}Se_{80}$ specimen.

Several parameters such as $C_p(T_g)$, transition width $\Delta T_g$, and the jump $\Delta C_p$ characterize the glass transition empirically. FWHM of the (α-relaxation) endothermic-peak roughly corresponds to the time-scale associated with the slow-relaxation of molecular entities in the glass. NRHF peak-position/kinetic glass-transition temperature is slightly lower in temperature than the caloric one ($T_k < T_g$) for most of the compositions (fig.2b). This suggests that as the temperature is increased the thermal energy first activates the diffusive (dissipative) motion of atomic-aggregates; further rise in its level enables excitation of a set of molecular vibrational states (reversible dynamics). The glass transition temperature $T_g(r)$ across the $Ge_xSe_{1-x}$ series is shown in the lower-right inset of fig.2b, along with the theoretical prediction by Tanaka [18] of its empirical behaviour, primarily for the chalcogenide glasses; $\ln T_g \cong 1.6\,r + 2.3$. A kink in $T_g$ is observed at $r = 2.4$, a maximum at $r = 2.74$, followed by the decrease indicating reduced vitreousity. At $x = 0$, the glass consists of long Se-chains, held together by the weak Van-der-Walls forces. Little thermal energy needed to convert it into the liquidus state means lower $T_g$ for small $x$'s. Relaxation timescale $\tau$ of a glass depends on the local structure, which goes up upon cooling, and diverges by the Kauzmann temperature $T_0$ [19]. Thus, a higher value of the α-relaxation time at $T_g$ denotes a relatively lower fictive temperature ($T_F$) of the glass-specimen and therefore, the existence of larger inhomogeneity.

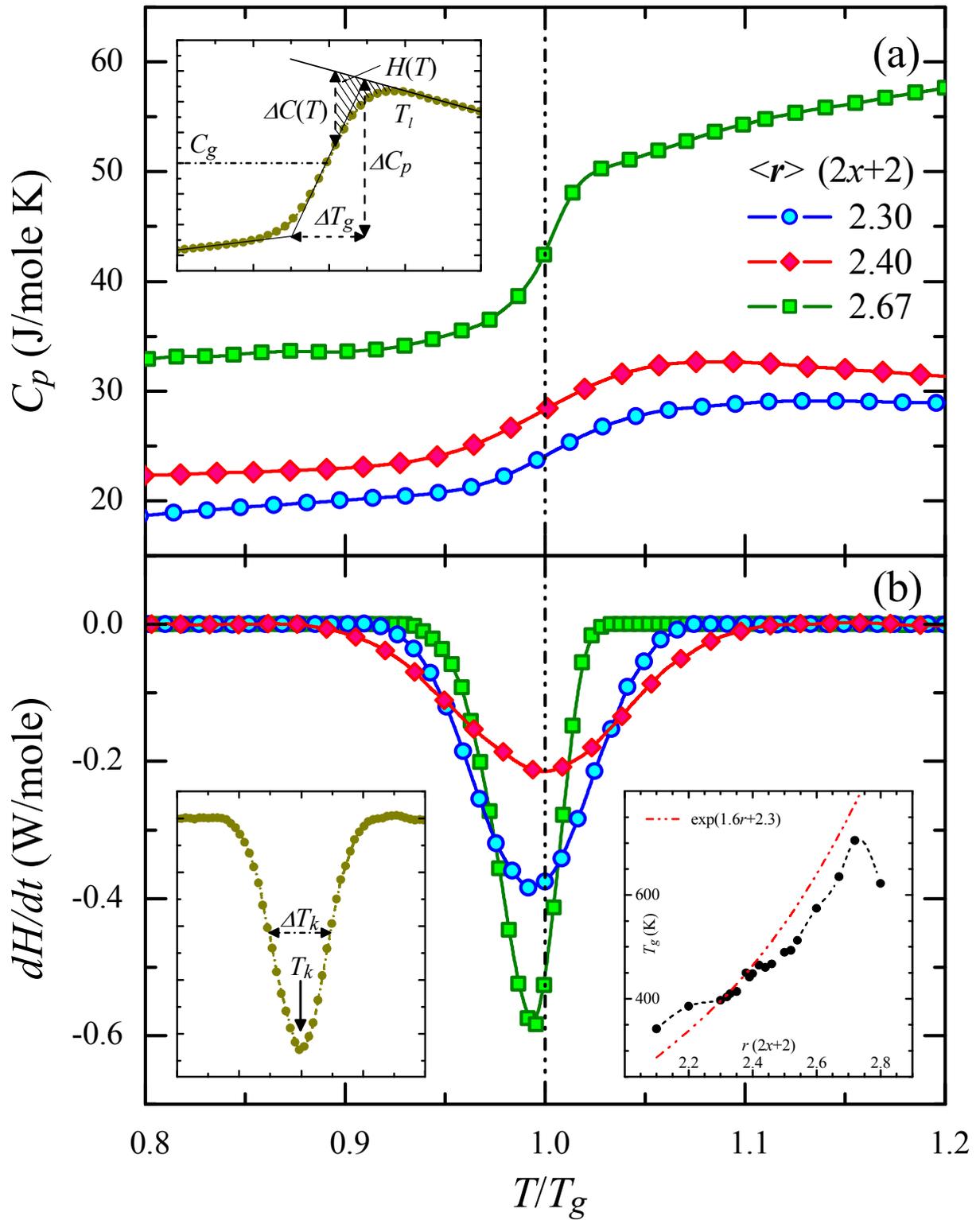

**Fig.2** Specific heat and α-relaxation peak signals vs. temperature (scaled to $T_g$). Insets show evaluation of the relevant parameters and the $T_g(r)$.

Adam-Gibbs formulation [20] relates the entropy and timescale of glass-formers as $\tau = \tau_0 \exp(C/TS_c)$. Vitreousity is robust and facilitated by *lesser difference* between thermodynamic-attributes of the liquidus and solidus phases [21]; *configurational* parts $\Delta C_p$ and $S_c$ of its specific heat and entropy respectively, which are proportionally-related [20]. Strong glasses typically feature smaller $\Delta C_p$ and $S_c$, and therefore longer $\tau(T_g)$ [21]. This is corroborated by the *reversibility-window* (R-W) [22] recognized in the network-glass families; ranges of <r> which feature nearly-reversible glass-transitions (negligible NRHF vs. rest of the family, due to extremely-slow relaxations over the R-W compositions). A glass-former with its metastable energy-landscape defined by configurationally-distant yet rather-degenerate local minima [23] explores its phase-subspace much too sluggishly; at high energy-cost of meandering through them. Such highly-frustrated *strong* glass-formers are little-susceptible to devitrification. The α-relaxation is controlled by two factors; (*a*) distribution of rigid molecular units within the clusters (formed already in the molten state) that undergo slow diffusion and internal rearrangement in time because of the 'free-volume' available in the glasses [20] and, (*b*) the size-evolution of these "cooperatively rearranging regions" (CRR) with cooling [24].

The typical non-exponential (KWW) decay of enthalpy $\{H = H(0)\exp[-(t/\tau)^\beta]\}$ in glasses characterizes *isothermal* relaxation in the *time-domain*, with no particular *experimental* timescale. On the other hand, DSC experiments carried out under *constant* temperature-ramp are *selective* in terms of the measurement-timescale ($\Delta t_{meas}$, determined via inverse of the ramp-rate $q$); windowing the signal as a band-pass filter. Moreover, the non-reversing heat flow NRHF($T$) ($\equiv \dot{H}$) happens to be the *initial* (i.e., near '$t$' ≈ 0 of sampling $T$ and averaged over $\Delta t_{meas}$) decay-rate of the yet-unrelaxed enthalpy $H_{ur}(T)$. Further, KWW stretched-exponential is but a convolution (quasi-continuous series-summation) of pure-Debyeans $\{\exp[-(t/\tau_{eff})^\beta] = \Sigma_i \exp[-(t/\tau_i)]\}$; individual $\tau_i$ defining a (rather-continuous) distribution $g_T(\tau_i)$, having a mean/median at (VFT) $\tau_{eff}(T) \equiv \tau_0\exp[E/(T-T_0)]$. Overwhelming contribution to $\dot{H}$ measured in the MDSC-run is due to the relaxation-process satisfying $\tau_i(T)=\Delta t_{meas}(q)$ *matched-filter* [25] condition at *each temperature*; reckoning with *reversely* (~VFT) $\tau_{eff}(T)$-shifting $g_T(\tau_i)$ that concurs the *q*-ramped *T*-scan, while the (removable) contributions from the *unmatched-timescales* manifest as sharper-noise and smoother-baseline.

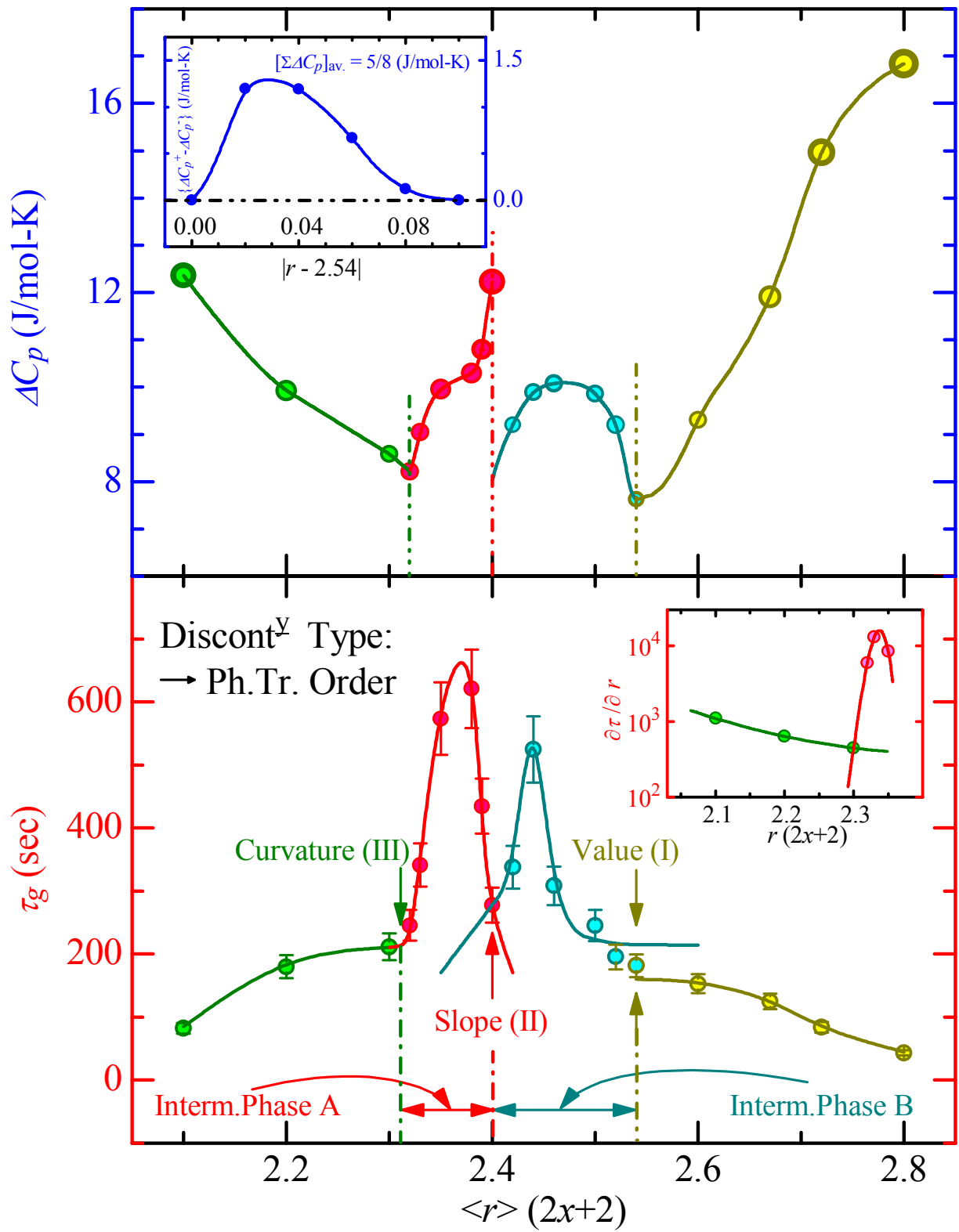

**Fig.3** Heat capacity jump and enthalpy-relaxation timescale in $Ge_xSe_{1-x}$ glasses.

The processed NRHF $\{\dot{H}(T)\}$ analysed here is essentially the ($g_T(\tau_i)$-weighted/$\Delta t_{meas}$-averaged) '*zero-time-derivative*' of the $\Delta t_{meas}(q)$-*compatible* Debyean (purely-exponential) term in the summed-series. At $T_g(q)$ (NRHF-maximum) the $\tau_i(T_g) = \Delta t_{meas}(q) = \tau_{eff}(T_g)$ *resonance condition* holds. For the MDSC data therefore, $\text{NRHF}(T_g) = \dot{H}_{max} = -[H_{ur}/\tau_{eff}]_{T_g}$ is accurately applicable. Ratio of the *configurational* enthalpy $H(T_g) = \int_{T_g}^{T_l}[C_l(T) - C_p(T)]dT$ (shaded area in fig.2a inset) to its loss-rate $\dot{H}(T_g)$ (NRHF ordinate, fig.2b) thus provides the α-relaxation time $\tau_{eff}(T_g) = \left|\frac{-1}{q}\frac{H(T)}{dH/dT}\right|_{T_g}$, using the temperature-time equivalence [26]. Configurational part of the heat capacity ($C_{\text{conf}}$) at a temperature $T$ is obtained here as the difference of $C_p$ from the linearly-extrapolated $C_l$ in the liquidus state.

Series-behaviors of heat capacity jump $\Delta C_p$ and α-relaxation time $\tau_g$ at $T_g$ are shown in fig.3, and exhibit *lower jumps* and *longer times* for the 'intermediate' compositions, regarded as more inhomogeneous; consistent with their higher glass-forming ability (GFA) and thermo-mechanical stability [27]. One cannot, however help noticing the *anticlimactic anomalous excursion* right at the celebrated threshold $r_p$ = 2.4; standing out as an exception to the general anticipation. Generically, characteristic time for mechanical relaxations is ~100 sec [16], and for structural relaxations on the order of 200sec [20], while a significant part of structural changes occurs after 400sec [17]. Recognizing that both Ln($\tau$) and the configurational entropy $S_c$ have identical (VFT) temperature dependence [20, 28], we suggest the following ansatz. Consider Ge$_x$Se$_{1-x}$-series as a '*thermodynamic-system*', with mean-coordination $r$ =2x+2 (or its functional) representing an '*analogous temperature*'; $\tau_g(r)$ and $\Delta C_p(r)$ then represent metrices of its (configurational) entropy and heat capacity respectively. Further, they are thermodynamic-derivatives of some (configurational) free-energy [27, 29] $F(r)$; so that $\tau_g(r) \stackrel{\text{def}}{=} dF(r)/dr$ and $\Delta C_p(r) \stackrel{\text{def}}{=} d^2F(r)/dr^2$. It now becomes obvious as to why & how the singularities in $\Delta C_p(r)$ and $\tau_g(r)$ correspond so well, consistent with the order of compositional '*phase-transitions*' recognized at $r_c$=2.31, 2.4, & 2.54.

A striking illustration of this consistency is the 2$^{nd}$-order floppy to isostatic transition at $r_2$ = 2.4, registered as a *slope-jump* in $\tau_g(r_2)$ (read $S_c \stackrel{\text{def}}{=} dF/dr$, [28]) and equivalently, as a *value-jump* in $\Delta C_p(r_2)$ (discontinuous 2$^{nd}$-derivative $d^2F/dr^2$). As a confirmation, the 1$^{st}$-order isostatic to stressed-rigid transition at $r_1 \approx 2.54$ ought to relate the *value-jump* in $\tau_g(r_1)$ ($S_c$, as discontinuous

1st-derivative $dF/dr$) to a corresponding '*hysteresis*' expected in (configurational) heat-capacity. Considering approach to $r_1$ from either side as '*cooling*' (the 'characteristic temperature' represented by $r_1$ is apparently *lower* vs. the neighborhood, as $\Delta C_p$ is locally-minimum there), a typical '*thermal cycling*' from $T_0(r_1)+\Delta T$ to $T_0(r_1)$ (say from $r_1-\delta$ to $r_1$) and back to $T_0(r_1)+\Delta T$ (from $r_1$ to $r_1+\delta$ say) picks a change in the (configurational) heat-capacity; $\Delta C_p^{hys}(\delta) = \{\Delta C_p(2.54-\delta) - \Delta C_p(2.54+\delta)\} \neq 0$. Inset in fig.3a shows this "*hysteretic-difference vs. the coverage ($\delta$)*", obtained utilizing the spline-curve fitted onto the experimentally determined $\Delta C_p(r)$. The average obtained by integrating this '*hysteresis-function*' $\Delta C_p^{hys}(\delta)$ is ~ 5/8 J/mol-K; which may signify the '*latent heat*' associated with the *first-order transition* at $r_1$ = 2.54. An intriguing novel outcome from close examination of the features is the slope-jump in $\Delta C_p(2.31)$ (of $d^3F/dr^3=d^2/dr^2\{dF/dr\}$), with a corresponding discontinuity expected in $d^2\tau_g/dr^2$ near $r_3$ ~ 2.3. Inset of fig.3b shows the *slope-points $d\tau_g/dr$*, obtained by differentiating the separately-fitted spline curves onto $\tau_g(r)$ (three points $\leq r_3$ and four points $\geq r_3$). Polynomials fitted onto these two sets of *derived $d\tau_g/dr$*-points join at $r_3$ = 2.31 with different slopes; consistently confirming the curvature-jump expected in the corresponding configurational entropy $S_c(r)$. While the 1st-order ($r_1$) and 2nd-order ($r_2$) transitions and the intermediate-phase/reversibility-window are well investigated by spectroscopic/thermal [21, 22] and structural [10, 29] studies, the 3rd-order transition presented here is witnessed for the first time.

## 4. Conclusions

At the glass transition temperature the $\alpha$-relaxation times in undercooled liquids are too long for the *macroscopic* changes to occur during experimental times. Providing the crucial thermodynamic evidence, our caloric and enthalpic-relaxation timescale studies clearly discern the rigidity-transitions that are well-recognized via vibrational and spectroscopic investigations. The present results facilitate narrowing-down the options for theoretical developments and open up further research for other network glasses. It would be interesting to explore the generality of the novel rigidity-transition found here at $r_3$ = 2.31 and its connection to the sole 3rd-order transition originally predicted at $r_p$ = 2.4 by the mean-field constraint-counting theories [3, 30]; quite suppressed to a lower-$r$ by the opening of the intermediate phase, as strong influence of the fluctuation-effects.